\documentclass[twocolumn]{aastex62}

\usepackage{color}
\usepackage{graphicx}	
\usepackage{amsmath}	
\usepackage{amssymb}	

\newcommand{\medd}{\dot{m}_\mathrm{Edd}}

\newcommand{\Qb}{Q_\mathrm{b}}
\newcommand{\mevpnuc}{\text{MeV} / \text{nuc}}
\newcommand{\unity}{\text{g cm$^{-2}$}}
\newcommand{\unitF}{\text{ergs cm$^{-2}$ s$^{-1}$}}

\received{XXXX}
\revised{XXXX}
\accepted{XXXX}
\submitjournal{ApJ}

\shorttitle{Burning physics and burst oscillations}
\shortauthors{Chambers et al.}

\begin{document}

\title{Burning in the tail: implications for a burst oscillation model}

\correspondingauthor{F. R. N. Chambers}
\email{frnchambers@uva.nl}

\newcommand{\API}{Anton Pannekoek Institute for Astronomy, University of Amsterdam, Postbus 94249, 1090 GE Amsterdam, The Netherlands}

\author{Frank R.N. Chambers}
\affil{\API}

\author{Anna L. Watts}
\affil{\API}

\author{Laurens Keek}
\affil{Department of Astronomy, University of Maryland, College Park, MD 20742, USA}

\author{Yuri Cavecchi}
\affil{Mathematical Sciences and STAG Research Centre, University of Southampton, SO17 1BJ, UK}
\affil{Department of Astrophysical Sciences, Princeton University, Peyton Hall, Princeton, NJ 08544, USA}

\author{Ferran Garcia}
\affil{Department of Magnetohydrodynamics Helmholtz-Zentrum Dresden-Rossendorf, POB 51 01 19, 01314 Dresden, Germany}
\affil{\API}

\begin{abstract}
Accreting neutron stars (NS) can exhibit high-frequency modulations, known as burst oscillations, in their lightcurves during thermonuclear X-ray bursts. Their frequencies can be offset from the spin frequency of the NS (known independently) by several Hz, and can drift by 1-3 Hz. One plausible explanation for this phenomenon is that a wave is present in the bursting ocean that decreases in frequency (in the rotating frame) as the burst cools. The strongest candidate is the buoyant $r$-mode; however, models for the burning ocean background used in previous studies over-predict frequency drifts by several Hz. Using new background models (which include shallow heating, and burning in the tail of the burst) the evolution of the buoyant $r$-mode is calculated. The resulting frequency drifts are smaller, in line with observations. This illustrates the importance of accounting for the detailed nuclear physics in these bursts.
\end{abstract}



\section{Introduction}
\label{sec:intro}
Type-I bursts are caused by runaway thermonuclear burning in the ocean layer of a NS; depending on the fuel available at the ignition site, the physics of bursts can vary greatly \citep{Galloway17}. This fuel is influenced by a number of factors: the material accreted from the donor star; the rate at which this material is accreted; the gravity at the surface; the ashes from previous bursts; and an extra source of heat from the outer crust known as \textit{shallow heating}. This mysterious heat source has been suggested as a resolution to a number of puzzles: the temperature evolution of some transient NS as they cool once accretion has ceased (see for example \citealt{Brown09,Deibel15,Degenaar15,Turlione15,Wijnands17}); the possible need to move superburst ignition depths to lower column depth to explain recurrence times and energetics (\citealt{Cumming06,Keek11}; although for an alternative  resolution see \citealt{Tumino18}); and transitions between different burning regimes \citep{intzand12,Linares12}. Most recently, \citet{Keek17} (KH17) showed that higher deep ocean temperatures could also explain the occurrence of Short Waiting Time (SWT) bursts at the accretion rates observed. There are a number of mechanisms that could lead to heating in the crust that would affect the ocean: pycno-nuclear and electron capture reactions in the crust generate a heat flux into the NS envelope \citep{Haensel90,Haensel03,Gupta07}. However, the postulated shallow heating would need to be something additional on top of this; its nature remains unclear \citep[see discussion in][]{Deibel15}. An additional heat source of this kind would have broader implications on other phenomena exhibited by NSs; this paper is particularly interested in the unsolved problem of \textit{burst oscillations}.

Timing analysis of some Type-I X-ray bursts reveals periodic oscillations throughout the lightcurve. These burst oscillations arise from asymmetries in surface brightness; however, the underlying mechanism responsible  has yet to be identified. The observed frequencies are either at, or offset by $\sim 3$ Hz from, the NS spin frequency (known independently for some stars; for a review, see \citealt{Watts12}), and may drift by $1-3$ Hz during the burst\footnote{Drifts are sometimes observed during the rising phase of bursts on accretion-powered pulsars with burst oscillations rather than the tail \citep[see e.g.][]{Chakrabarty03,Altamirano10a} but the properties of those burst oscillations are somewhat different from the non-pulsars \citep{Watts12}}. One possible explanation is the presence of ocean modes which would give rise to large-scale patterns, the drift speed changing during the burst as a result of the ocean cooling. These modes could plausibly be excited by bursts, which is why \citet{Heyl04} suggested them as a potential explanation for burst oscillations. The many different families of modes would lead to a variety of observable frequencies, which can be constrained based on observed properties of the oscillations. The best candidate is a low azimuthal wave number ($m$) buoyant $r$-mode driven by buoyancy in the ocean, and strongly affected by the Coriolis force.

\citet{Heyl04} assumed a simple two layer model to calculate the frequency of the mode which resulted in a larger drift than shown by observations \citep{Muno02a}.
\cite{Lee04} included a radial structure in the model for two envelope models, one convective and one radiative, meant to represent the ocean at the early and late stages of a burst respectively. They found that $r$-modes are driven unstable by nuclear burning in the convective zone and that, depending on the order of the mode and spin of the star, the frequency in the rotating frame was smaller by $2 - 10 \%$ in the radiative model than the convective model. Subsequently, \cite{Piro05b} (hereafter PB05) included a cooling model and calculated frequencies upon snapshots of an evolving ocean. The frequency drifts they found were too great (at $5$ Hz over $15$ secs), and so they suggested that the buoyant $r$-mode in the shallow bursting layer transitions to a crustal interface wave in the deep ocean during cooling in order to halt the drift and reduce surface modulations. This mechanism was shown not to be viable \citep{Berkhout08}, and so the model was to some degree disregarded.

An improved cooling model for the ocean would certainly affect the buoyant $r$-mode frequencies and drifts.
PB05 held the composition constant during the burst, assuming a two layer model with a single species in each layer, did not take into account nuclear burning throughout the burst and used a value for the heat flux in the outer crust that did not include the possible effect of shallow heating.

In this paper, in order to illustrate the importance all of these effects have on the mode calculation, we take snapshots from a model used to explain SWT bursts \citep{Keek17}, and calculate the frequency evolution of a buoyant $r$-mode.

\section{Frequency calculation}
\label{sec:T-evol-F-drift}
Modes that might explain burst oscillations are excited in the ocean of the NS; a thin layer of fluid composed of accreted H, He and some trace metals that burn to heavier elements as they sink deeper into the ocean. The material eventually reaches the crust, where heavy ions are bound in a lattice formation by Coulomb forces and surrounded by a sea of degenerate electrons \citep{Chamel08}. The phase transition between ocean and crust is defined as the point at which the ratio of Coulomb to thermal energy reaches: $\Gamma = 175$ \citep{Farouki93}.

Here we briefly outline how PB05 calculated mode frequencies during a burst \citep[for further discussion, see][]{Chambers18}. The mode equations are derived in spherically symmetric Newtonian gravity, by assuming adiabatic perturbations upon a thin static ocean layer on the surface of the star. The \textit{Traditional Approximation} simplifies the calculations significantly by neglecting the horizontal component of the rotation angular velocity vector, and results in a set of two ordinary differential equations which are separable in radial coordinate and latitude. The latitudinal component of the perturbation equations involves the operator defining \textit{Laplace's Tidal Equation}; an eigenvalue equation for shallow water waves $L_\mu f = - \lambda f$. The operator $L_\mu$ acts on the latitudinal component of the perturbation, and depends on $m$, the NS spin, and the mode frequencies $\omega$. Solutions of this equation are Hough functions \citep{Longuet-Higgins68}, and the solution with properties that best match the observational constraints of burst oscillations is a low $m$ buoyant $r$-mode \citep{Heyl04}.

We therefore study the $m=1$, $l=2$ (spherical harmonic degree\footnote{In the notation used by \cite{Lee97} this is the $k=-1$ $r$-mode.}) buoyant $r$-mode which has a strong maximum and thus high visibility. In general, the eigenvalue $\lambda$ depends on both the spin of the NS and the frequency of the mode through $q = 2 \Omega / \omega$. For the case of $r$-modes, however, for sufficiently large $q$ the eigenvalue becomes a constant value. Solutions to the mode equations are found by first choosing a wavevector based on the mode ($k^2 = \lambda / R^2$), and solving the radial equations using a shooting method, with the condition that perturbations are zero at the ocean-crust interface. Since the background evolves during cooling, various solutions are found using snapshots of this background. While the mode frequency is weakly sensitive to the inner boundary at the crust, it is sensitive to the outer boundary. The location of this boundary is fixed by the condition that the mode timescale is approximately equal to the thermal timescale (PB05), giving a column depth of $10^7$ $\unity$. This is also the location where the adiabatic condition is no longer valid.

Both temperature and composition affect the modes (through density gradients), with the frequency dependent on the difference in temperature between the bursting and cool layers. Using a simple two layer model, PB05 approximated the frequency of the mode as:
\begin{equation}
  \label{eq:freq-estimate}
  \frac{\omega / 2 \pi}{10.8 \text{ Hz}} =
  \left( \frac{\lambda}{0.11} \right)^{1/2}
  \left( \frac{10 \text{ km}}{R} \right)
  \left( \frac{T_b / \mu_b - T_c / \mu_c}{5 \times 10^8 \text{ K}} \right)^{1/2} ,
\end{equation}
where subscripts stand for \textit{bursting} and \textit{cool}, and $\mu$ stands for mean molecular mass per electron. The pressure in this estimate was derived assuming an ideal gas of electrons, neglecting degeneracy and radiation pressure. Degeneracy is significant in the cool layer, but lifted in the bursting layer where radiation pressure could also play an important role. From this estimate, one would expect that a slow rate of cooling would reduce frequency drift, and extra sources of heat from the crust would reduce offset from the spin frequency.

\section{Thermal evolution}
\label{sec:T-evol-F-drift}
Here we summarise the models for background cooling used in PB05 and KH17. These models differ in two ways: the presence of ongoing nuclear burning and the parameters which set composition and temperature profiles. In order to make clear the effects on mode frequencies of these two differences, we also calculate a new cooling model which takes the thermal evolution scheme from PB05 and parameters for composition and temperature profile from KH17; we refer to this model as PTKP (PB05 temperature, KH17 parameters) for the remainder of the paper.

Previously, PB05 approximated the temperature evolution during a burst by dividing the NS ocean into two layers: a \textit{hot layer} in which heat from nuclear burning is deposited in the form of an enhanced heat flux; and a cool \textit{base layer} with a persistent flux dictated by the crust. Heat from the hot layer can radiate from the surface at the outer boundary and conduct into the cool layer according to simple thermal diffusion. Nuclear energy generated throughout the burst is not taken into account, and the composition is fixed in each layer to post-burst ashes \citep[inspired by][]{Schatz01,Woosley04}. The key parameters that define this cooling model are the compositions and initial fluxes in each layer; PB05 tested three models (see Table 1, Section 3). The model they found with the smallest frequency drift (model 1 in their paper) consisted of $^{40}$Ca ($^{64}$Zn) in the bursting (cool) layer, and an initial condition of flux $10^{25}$ $\unitF$ ($10^{21}$ $\unitF$) in the bursting (cool) layer; these parameters match a system with an accretion rate of 0.1 $\medd$ and a base heating of 0.1 $\mevpnuc$.

The thermal evolution models in KH17 used the stellar evolution code \texttt{KEPLER} \citep{Weaver78,Woosley04}. This code employs an adaptive one-dimensional Lagrangian grid, and a large adaptive network of isotopes to follow nuclear burning. Chemical mixing between zones (due to convective processes) is approximated using mixing-length theory. The layer is initialised as an iron substrate in the base, and accretes lighter elements to make fuel for unstable burning.

In these simulations, the rp-process plays an important role in the nuclear burning \citep{Wallace81,Schatz01}. It consists of a fast and slow part: the fast part consists of the reactions at the start of the burst up to the first $\beta$-decay waiting points; and the slow part consists of the proton-capture and beta-decay reactions in the tail of the burst which are delayed by these waiting points \citep{Woosley04,Fisker08}. In the first burst in the sequence, the slow part of the rp-process dominates the lightcurve after 20 seconds. For the second and third bursts, ignition occurs when H mixes into the ashes, which are rich in rp-process seed nuclei. Upon ignition the protons are captured quickly, and there are not many left to continue the rp-process and power a long tail.

The cooling model is taken from a set of simulations outlined in KH17, which were used to investigate SWT bursts; we take the burst triplet found in their simulation\footnote{SWT bursts of this type have been observed from 15 sources \citep{Keek10}.}. The accretion rate in this cooling model was $0.1$ $\medd$, accreting approximately solar composition (mass fractions of 0.71 $^{1}$H, 0.27 $^{4}$He, and 0.02 $^{14}$N), and with high base heating of $3$ $\mevpnuc$. The luminosity and composition (about the ignition depth) of these three bursts is plotted in Figures 8 and 10 of KH17.

The new background model, PTKP, uses the thermal evolution scheme from PB05 (that is to say, with no heating from ongoing nuclear burning) and parameters for composition and temperature profile from KH17. The initial flux is the same as PB05 in the bursting layer at $10^{25}$ $\unitF$, but a much higher $4 \times 10^{22}$ $\unitF$ in the cool layer to match the high $\Qb$ used in KH17. The composition used in the bursting layer is 0.7 $^1$H, 0.24 $^{4}$He, and the remaining mass in equal quantity $^{12}$C, $^{14}$Ni, and $^{15}$O. The composition of the cool layer is changed to pure $^{56}$Fe, to match KH17. One extra difference between PB05 and PTKP is that the depth of composition and flux change are no longer the same. Composition changes at a column depth of $5 \times 10^7$ $\unity$, whereas the flux changes at $3 \times 10^8$ $\unity$. This is done to better match the density profile of KH17 where the ashes of previous bursts extend to shallower depths above the ignition location.

\begin{figure}
  \centering
  \input{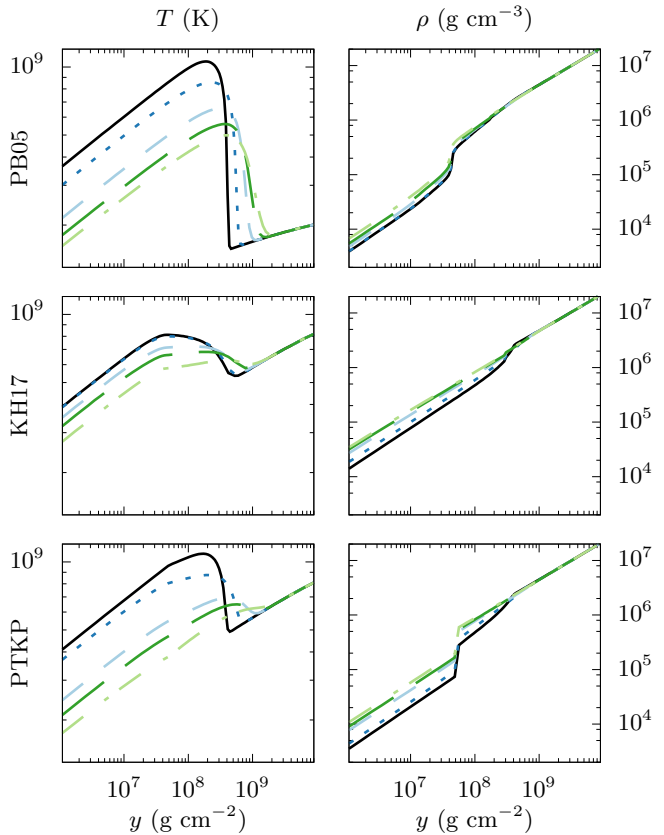}
  \caption{Temperature and density evolution of background cooling models tested in this work. The left column is temperature and the right column is density. The first row the model of PB05, the middle row the first burst from the triplet calculated in KH17 and right row PTKP (a cooling model using the thermal evolution scheme of PB05, and parameters to match the temperature and composition of KH17).
    Each line is a different time with solid (black), dots (dark blue), short-dash (light-blue), long-dash (dark-green), and dot dash (light-green) at $0.15, 1, 5, 10,$ and $20$ secs. The temperature discontinuity occurs at the ignition site of $3 \times 10^8$ $\unity$, while the change in density for KH17 and PTKP occurs at a shallower depth of $5 \times 10^7$ $\unity$ where the composition changes. The KH17 model includes burning in the tail of the burst, while PB05 and PTKP use simple heat diffusion.}
    \label{fig:temp-evol}
\end{figure}

Fig. \ref{fig:temp-evol} plots the temperature and density of: a reproduction of PB05, the first burst from KH17, and PTKP for several time steps after the peak. The background in the cool ocean layer matches well between the KH17 and PTKP while at the ignition site the temperature profiles of these two models are not well matched in shape. PB05 and PTKP exhibit much more rapid cooling in the bursting layer; it is reasonable to expect that this faster changing temperature will result in a larger frequency drift.

\section{Results}
\label{sec:results}
From the point of view of how these cooling models affect the mode frequency, we note the following difference given the estimate in eq. \ref{eq:freq-estimate}. The temperature in the cool layer is significantly higher in KH17 and PTKP at $\sim 7 \times 10^8$ K (compared to PB05 at $2 \times 10^8$ K) which should reduce absolute frequencies (in the rotating frame). A smaller mean molecular weight per electron in the bursting layer is expected to increase absolute frequencies; KH17 and PTKP both contain a large fraction of Hydrogen in their bursting layer (with $\mu_{\text{b}} = 1.18$, compared to $\mu_{\text{b}} = 2$ in PB05).

Inspecting the temperature evolution in the bursting layer, PTKP is no faster at cooling than PB05 -- if anything it is slower; the peak temperature at the beginning for both models is $10^9$ K, and at $10$ seconds is $5 \times 10^8$ K for PB05 and $6 \times 10^8$ K for PTKP. However, PTKP has a higher temperature in the cool layer making the difference in temperature between the two layers much smaller than for PB05. Using these parameters the temperature and composition dependent factor in eq. \ref{eq:freq-estimate} for the two models, $ \left( T_b / \mu_b - T_c / \mu_c \right) / 5 \times 10^8 $ K, at the start of cooling is $0.8$ for PB05 and $0.99$ for PTKP, and at $10$ seconds $0.3$ for PB05 and $0.32$ for PTKP. From this result it should be expected that PTKP has a higher frequency than PB05, and a greater drift. The same factor for KH17 goes from $0.75$ to $0.42$, implying smaller frequencies and frequency drifts.

\begin{figure}
  \centering
  \input{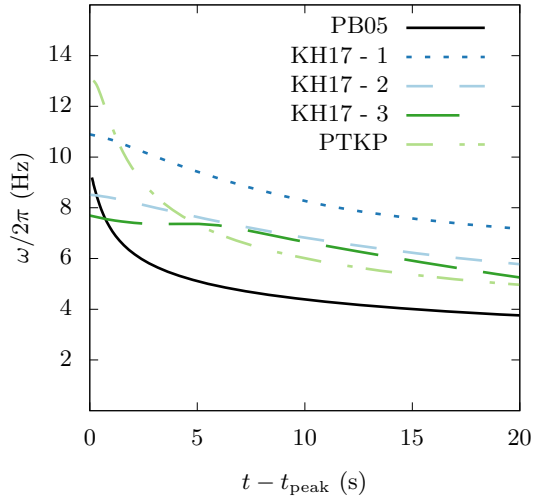}
  \caption{The results of the frequency evolution of the $m=1$, $l=2$ buoyant $r$-mode for the various cooling models outlined in this letter. Plotted are the frequency evolution of PB05 (included for reference), each burst in the KH17 triplet and a model which combines the thermal evolution scheme of PB05 with the burst environment calculated in KH17 (PTKP, see text).}
  \label{fig:frequencies}
\end{figure}

Fig. \ref{fig:frequencies} plots the frequencies of the $m=1$, $l=2$ buoyant $r$-mode with a single radial node for PB05, each burst in the triplet calculated by KH17, and PTKP. The frequency drift for each burst in KH17 is approximately the same at $2 - 3$ Hz over $15$ seconds, with the first burst marginally greater. The third burst exhibits a plateau in frequency between $3$ and $5$ seconds which matches the second peak in luminosity and is due to extra burning processes depositing heat in the layer after ignition. The frequency drift for PTKP is more than twice as great as KH17 at $\sim 8$ Hz over $15$ seconds. The initial frequency is also $2$ Hz higher due to a higher peak temperature. The frequencies of KH17 and PTKP match at around $1$ sec after the peak, which is to be expected as at this point the temperature profiles match the most closely (single-dot dash lines in fig. \ref{fig:temp-evol}). Compared to PB05, rotating frame frequencies calculated in KH17 and PTKP are higher due mostly to there being a lighter ocean. The frequency drift is overall greater in the models without burning in the tail of the burst due to the faster cooling rate.

\section{Discussion}
\label{sec:discussion}
The new results for frequency drift are quite different to those published previously. The drifts in PB05 were $>6$ Hz over $20$ seconds, while for KH17 (the model that includes ongoing nuclear burning) drifts are at most $4$ Hz. This change is not due only to different burst conditions, but mostly a result of the presence of ongoing nuclear burning in the tail of the burst as demonstrated by the frequency drift in PTKP of $8$ Hz over $20$ seconds, where burning is absent. This point demonstrates that when predicting the frequency drift it is crucial to accurately model the reactions ongoing in the burst that set the temperature of the layer \citep[for example the rp-process][]{Cyburt16,Ong17}.

The extra Hydrogen fraction present in these bursts raises an interesting issue, since \citet{Cumming00} showed that a shearing layer could act to wash out oscillations from deeper regions propagating to the surface - a problem particularly significant for mixed H/He bursts. However, these results were more problematic for backgrounds with a higher temperature, where the luminosity was a significant fraction of the Eddington luminosity, because the thermal timescale tends to increase with temperature and radiation pressure becomes more important. The bursts examined in this paper are quite weak in comparison. More investigation is required into how a large accreted Hydrogen fraction would affect the amplitude and visibility of burst oscillations.

It is also uncertain whether observational evidence supports this picture, since although oscillations have not been observed from the canonical Hydrogen-rich long rp-process tail burster GS 1826-24 \citep[see for example][]{Heger07b} burst oscillations are observed during mixed H/He bursts.  Moreover, the sequence of bursts from 4U 1636-53 \citep{Keek10} modelled by KH17, for which Hydrogen does play a role in the bursting layer (see Section \ref{sec:T-evol-F-drift}), does have detectable burst oscillations in the third burst in the sequence \citep{Bilous18}.

Some immediate questions for this model include testing a wider variety of burst scenarios with different accreted compositions, accretion rates, and base heating. This testing should include thorough comparison against the data \citep[for tests against accretion rate, see][]{Franco01,Muno04,Ootes17}. In particular, it would be interesting to study the detectability of burst oscillations as a function of hydrogen content in the burning layer. Other interesting questions involve changing the type of mode, or wave vector $k^2 = \lambda / R^2$, which acts to dramatically decrease rotating frame frequencies, reducing the offset from the spin frequency as seen by an inertial observer.

Extra physics in the ocean, like chemical separation, would affect the ocean-crust boundary as heavy nuclei freeze out more easily compared to light nuclei. Changing the location of this transition would have a small effect on frequencies. However, if the background conditions were to be significantly altered (through extra heat, temperature changes, or composition changes) frequencies would certainly be altered.

The burst oscillation frequency calculated here has a greater offset from spin than previous models which suggests that if this model is correct, the spin frequency inferred from burst oscillations might be larger than previously thought (for sources with no independent confirmation via accretion-powered pulsations). The degree to which frequencies would be offset from the spin (although not the amount of frequency drift) will also be affected by relativistic effects, which are estimated to lead to a reduction of up to $20 \%$ in the rotating frame \citep{Maniopoulou04}. A change in offset from the spin frequency would have implications for efforts to infer equation of state or mass and radius from pulse profile modelling of burst oscillations, since NS spin is an important element of the space-time model \citep[see for example][]{Riley18}. It would also be important for continuous gravitational wave searches from accreting NS \citep[see for example][]{Watts08a}.

In summary, adding more accurate physics to the background has implications for burst oscillations. Heat from nuclear reactions in the tail of the burst helps to slow cooling and therefore reduce frequency drift, which reinstates the buoyant $r$-mode model as a viable candidate for the non-pulsars. The previous models of \cite{Heyl04} and PB05 over-predicted the drift, these new models are more in line with observations. The burst oscillation mechanism is still not solved, but this study does demonstrate the importance of having realistic burst models that include ongoing nuclear burning, and the broader implications of shallow heating.

\section*{Acknowledgements}

The authors would like to thank the referee for their helpful comments, particularly their input on the potential role of Hydrogen in these bursts.
FRNC, ALW and FG acknowledge support from ERC Starting Grant No. 639217 CSINEUTRONSTAR (PI Watts). FG was supported by a postdoctoral fellowship of the Alexander von Humboldt Foundation. YC is supported by the European Union Horizon 2020 research and innovation programme under the Marie Sklodowska-Curie Global Fellowship grant agreement No 703916. This work benefited from discussions at the BERN18 Workshop supported by the National Science Foundation under Grant No. PHY-1430152 (JINA Center for the Evolution of the Elements).


\bibliographystyle{aasjournal}

\end{document}